\documentclass[aps,prl,amsmath,amssymb,twocolumn,superscriptaddress]{revtex4-1}  
\usepackage{graphicx}  
\usepackage{color}
\usepackage{amssymb}   
\usepackage{amsmath}
\usepackage{epstopdf}
\usepackage{natbib}
\usepackage{hyperref}  
\definecolor{myblue}{RGB}{46, 48,146}
\hypersetup{colorlinks=true,linkcolor=myblue,citecolor=myblue,urlcolor=myblue, linktocpage}
\usepackage{bm}		   
\usepackage{braket}
\usepackage{gensymb}
\usepackage{bbold}
\usepackage{comment}
\usepackage{xcolor}

\begin{document}

\title{Colossal Layer Nernst Effect in Twisted Moir\'{e} Layers}
    
\author{Jin-Xin Hu}

\affiliation{Division of Physics and Applied Physics, School of Physical and Mathematical Sciences, Nanyang Technological University, Singapore 637371} 

\author{Chuanchang Zeng}
\thanks{zengcc@baqis.ac.cn}
\affiliation{Beijing Academy of Quantum Information Sciences, Beijing,100193, China} 	

\author{Yugui Yao}
\thanks{ygyao@bit.edu.cn}
\affiliation{Centre for Quantum Physics, Key Laboratory of Advanced Optoelectronic Quantum Architecture and Measurement(MOE),
School of Physics, Beijing Institute of Technology, Beijing, 100081, China}
\affiliation{Beijing Key Lab of Nanophotonics $\&$ Ultrafine Optoelectronic Systems,
School of Physics, Beijing Institute of Technology, Beijing, 100081, China}

\date{\today}

\begin{abstract}
In this work, we establish a theoretical analysis of the emergence of layer-contrasted Nernst response perpendicular to the direction of the temperature gradient in twisted moir\'{e} layers, called layer Nernst effect (LNE). This phenomenon arises from the trigonal warping of the Fermi surface along with a layer-contrasted pseudomagnetic field. Interestingly, the Fermi surface's warping explicitly breaks intra-valley inversion symmetry, which leads to an imbalance between left- and right-moving carriers, thus resulting in a non-vanishing LNE. We then validate our theoretical scheme by applying it to twisted bilayer graphene (TBG). Importantly, we find that the LNE coefficient in TBG can reach values as high as $10^3$~A/(m$\cdot$K), surpassing those of previously known materials by at least one order of magnitude. These results provide a theoretical foundation for utilizing TBG and other twisted moir\'{e} layers as promising platforms to explore layer caloritronics and develop thermoelectric devices.
    \end{abstract}
\maketitle
\emph{Introduction.}---Nernst effects are of essential importance in realizing the coordinated control of heat and charge in modern electronics. Primarily, the Nernst effect is characterized by the generation of an electric signal transverse to the temperature gradient under a perpendicular external magnetic field~\cite{checkelsky2009thermopower,behnia2016nernst}. 
Attributed to the endowed crucial properties of Berry curvature, anomalous Nernst effect~\cite{xiao2006berry,miyasato2007crossover,sakai2018giant,weischenberg2013scattering,cheng2008spin} and the cousin versions in terms of the spin and valley degree of freedom~\cite{meyer2017observation,cheng2008spin,tauber2012extrinsic,yu2015thermally,sharma2018tunable,dau2019valley} have been largely studied. The linear Nernst effect quantified by Nernst coefficient linear in the temperature gradient vanishes under time-reversal ($\mathcal{T}$) symmetry in general. Interestingly, recent studies have shown the nonlinear Nernst current, which arises as a second-order response to the applied temperature gradient, can survive in $\mathcal{T}$-invariant systems~\cite{yu2019topological,zeng2019nonlinear, Fu_2021_NLANE}.

\begin{figure}
		\centering
		\includegraphics[width=1\linewidth]{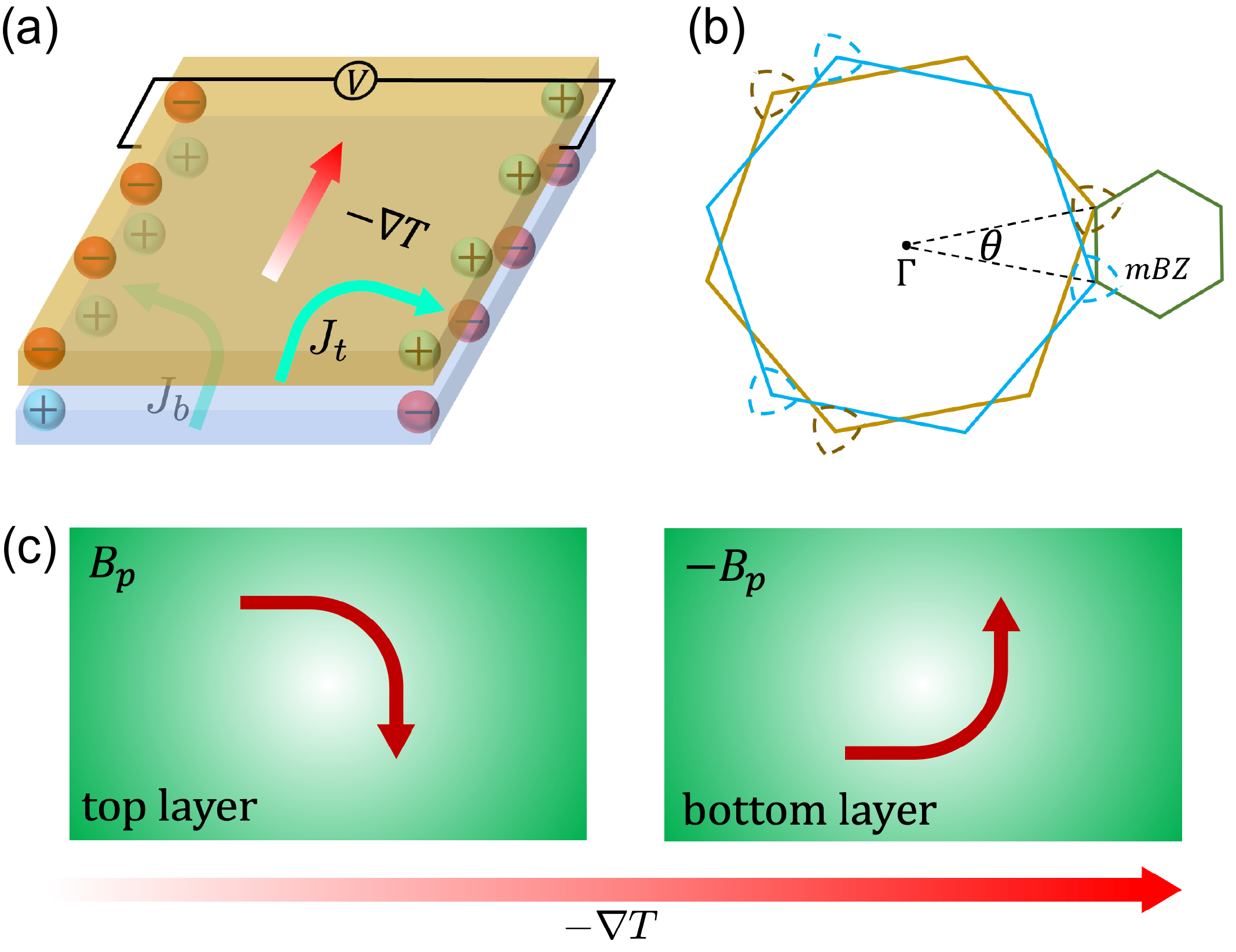}
		\caption{(a) Schematic illustration of the layer Nernst effect in a bilayer system. Under a temperature gradient $\nabla T$, layer-contrasted current flows, denoted as $J_t$ and $J_b$ for the top and bottom layers, are generated perpendicularly to $\nabla T$. (b) In the momentum space, the original Brillouin zone of the bottom (blue) and top (brown) layers is folded into moir\'{e} Brillouin zone (mBZ, green). The trigonally warped Fermi pockets are around the Brillouin zone corners. $\theta$ is the twist angle between the two layers. (c) The layer pseudomagnetic field ($\pm B_p$) drives the layer contrasted current flows under the $\nabla T$.}
		\label{fig:fig1}
\end{figure}

Despite the intense studies over the past few decades, the Nernst effect remains underutilized in practical applications~\cite{Behnia_2016_review, He_2017_review}. A small Nernst coefficient and/or thermal gradient would lead to minuscule Nernst signals, significantly obstructing the probing and potential applications. As a result, the Nernst effects of more evident response, namely, with a larger Nernst coefficient, are still in high demand. 

Long-period moir\'{e} materials formed in van der Waals heterostructures have evoked significant interest since the discovery of correlated insulator and superconductivity in twisted bilayer graphene (TBG) \cite{cao2018unconventional,cao2018correlated,yankowitz2019tuning,balents2020superconductivity}. Amounts of nontrivial topological properties, including the spontaneous ferromagnetism~\cite{sharpe2019emergent,serlin2020intrinsic} and various transport effects such as the magneto-electric and nonlinear Hall effects~\cite{he2020giant,zhang2022giant,pantaleon2021tunable,duan2022giant,hu2022nonlinear,sinha2022berry,huang2023giant,huang2023intrinsic}, have been demonstrated in the moir\'{e} systems. In addition, it was proposed recently that TBG can support the so-called layer Hall effect \cite{zhai2023time,zhu2023layer}, where layer-contrasted Hall current is generated due to interlayer hybridization, 
originating differently from that induced by layer-locked Berry curvature~\cite{gao2021layer}. In contrast, only a few studies so far have focused on the thermoelectric transports in TBG~\cite{ghawri2022breakdown,mahapatra2020misorientation,mahapatra2017seebeck,paul2022interaction}, significantly less investigated. Based on the Onsager reciprocity~\cite{LD_book}, the layer-contrasted Nernst response is expected to emerge naturally in these systems supporting nonzero Hall response.

In this work, we propose a new type of Nernst effect that offers a remarkable Nernst response, namely the layer Nernst effect (LNE) in the $\mathcal{T}$-invariant twisted moir\'{e} systems. Utilizing the Boltzmann transport equation approach, we establish the formalism for LNE induced by the layer velocity curvature (LVC). This is of a fundamentally different origin from all the previously reported Nersnt effects. We also explicitly demonstrate the significant role of breaking the intravalley inversion ($\mathcal{I}$) by the trigonal warping effect. Importantly, the LNE is forced to be zero if intravalley $\mathcal{I}$ is preserved when warping is absent. Moreover, we find that the LNE conductivity of TBG can reach magnitudes as high as approximately of the order in $0.1- 1~\mu$A/K (effectively $10^2 -10^3~$~A/(m$\cdot$K) for a layer device of nanometer-scale thickness), far better than the currently achieved large Nernst coefficients [$0.5-45~$~A/(m$\cdot$K)] that have been reported in only a few topological materials at similar temperatures~\cite{Ikhlas_2017_ANE, Sakai_2018_ANE, ANE_Li_2020PRM, ANE_Asaba_2021_SciAdv}. Finally, we extend our formalism of LNE to twisted multilayer graphene systems and also analyze the intriguing layer responses in twisted double-bilayer graphene (TDBG), a system that has been successfully fabricated in experiments recently~\cite{liu2020tunable,shen2020correlated}.

\emph{General formalism of LNE in bilayer structures.}---Conventionally, a current flow density can be obtained utilizing Boltzmann transport theory if the flowing carrier's velocity along with its momentum- or energy-dependent distribution function is known \cite{cercignani1988boltzmann}. Due to the layer-like spatial separation for the carriers in coupled bilayer systems, a layer-resolved velocity could be defined as $\hat{\bm{v}}_L=\{ \tau_z,\hat{\bm{v}}\}/2$, where $\hat{\bm{v}}$ is the normal velocity operator, and $\tau_z$ acts on the layer pseudospin space \cite{sinova2004universal,xu2014spin}. Following the conventional analogy, it leads us to a layer-resolved current density, which can be written as
\begin{equation}
\label{eq:currentflow}
\bm{J}_L=-e\int_{\bm{k}}f(\bm{k})\bm{v}_L(\bm{k}).
\end{equation}
Here, $\bm{v}_L(\bm{k})=\langle u(\bm{k})| \hat{\bm{v}}_L|u(\bm{k})\rangle$, $\int_{\bm{k}}=\int d^2\bm{k}/(2\pi)^2$, and the summation over the band $(|u(\bm{k})\rangle)$ index has been omitted. $f(\bm{k})$ is the Fermi-Dirac distribution of the carriers in the occupied states, which is apt to be perturbed by external fields, giving rise to various transport responses. Considering a homogeneous and uniform temperature gradient $\bm{\nabla}T$, its first-order deviation in $\bm{\nabla} T$ can be then obtained as $f^T_1=\tau\hbar^{-1}[\frac{\varepsilon(\bm{k})-\mu}{T}]\partial_{\alpha}f_0 \nabla_{\alpha}T$, with $\tau$ being the constant relaxation time 
and $\partial_{\alpha}=\partial/\partial_{k_{\alpha}}$. Such a temperature-driven deviation directly brings us to the layer-contrasted current flow, linear in and transverse to $\bm{\nabla} T$, which reads as 
\begin{equation}
\label{eq:layercurrent}
J_L^y=-\tau \frac{e}{T}\int_{\bm{k}}[\varepsilon(\bm{k})-\mu]f_0'v_x(\bm{k}) v_L^y(\bm{k})\nabla_x T.
\end{equation}
Here $f_0' = \partial f_0/\partial \varepsilon$ and $\alpha = x$ is considered for $\nabla_{\alpha}T$. Straightforwardly, the layer Nernst coefficient, defined as $\alpha_N^L= (\alpha^L_{xy}-\alpha^L_{yx})/2$, can be obtained as
\begin{equation}
\label{eq:Nernstcoiffe}
\alpha^L_{N}(\mu)=\tau\frac{e}{2T}\int_{\bm{k}}[\varepsilon(\bm{k})-\mu]\Omega_L(\bm{k})\delta_F[\varepsilon(\bm{k})-\mu]
\end{equation}
where $\Omega_L(\bm{k}) = \bm{v}(\bm{k})\times \bm{v}_L(\bm{k})$ is denoted as the LVC, $\bm{v}(\bm{k})=\hbar^{-1}\bm{\nabla}_{\bm{k}}\varepsilon(\bm{k})$, and $\delta_F[\varepsilon(\bm{k})-\mu]=-\partial f_0/\partial \varepsilon=\{4k_B T\cosh^2[(\varepsilon(\bm{k})-\mu)/2k_B T]\}^{-1}$ is the delta-like function. Note that LVC has the same physical origin as the layer current vorticity introduced in Ref.~\cite{zhai2023time}. The pure Nernst coefficient for each layer is then given by $\alpha_N^t=-\alpha_N^b=\alpha_N^L/2$, respectively corresponding to the LNE current $\bm{J}_{t/b}$ for the top/bottom layer, as schematically shown in Fig.~\ref{fig:fig1}~(a).

Intuitively, as long as the current in each layer remains different, the layer-resolved current naturally emerges in the bilayer systems, i.e., $\bm{J}_L=\bm{J}_t-\bm{J}_b \neq 0$, despite the detailed driven fields. Specifically, however, whether LNE survives depends on the symmetry constraints for the given bilayer system. One easily finds $\Omega_L(\bm{k})$ is intervalley $\mathcal{T}$-even but is intravalley $\mathcal{I}$ odd, which indicates that a $\mathcal{T}$ invariance and $\mathcal{I}$ breaking will be required to support the finite LNE response in bilayer systems. Following a similar analogy, the layer Hall conductivity $\sigma_H^L$ can also be obtained when electric field $E$, instead of $\nabla T$, is applied (see Ref.~\cite{zhai2023time} and Supplemental Material~\cite{NoteX}). Interestingly, the Mott relation is found to be well obeyed between $\sigma^L_{H}$ and $\alpha^L_{N}$ at low temperatures, i.e., 
\begin{equation}
\label{eq:Mottrelation}
\alpha^L_{N}(\mu)=\frac{\pi^2 k_B^2 T}{3e}\frac{\partial\sigma^L_{H}(\mu)}{\partial \mu},
\end{equation} 
which has been discussed extensively for the anomalous Hall and Nernst effects \cite{xiao2006berry}, and is valid only when the energy derivative of the electric coefficient is continuous and at relatively low temperatures ($k_B T\ll \mu$)~\cite{xiao2016unconventional}. We also want to mention that, in the second-order transport regime, the layer Hall and Nernst current is found to be $\propto \partial_{\bm{k}}\Omega_L$, which is $\mathcal{T}$-odd, rendering the net contributions from terms $\propto E^2$ and $\propto(\nabla T)^2$ be zero for $\mathcal{T}$-invariant systems. We can also expand them to higher order terms like the Berry curvature multiple in the recent studies~\cite{zhang2023higher,zeng2021BCD}.  

\begin{figure}
		\centering
		\includegraphics[width=1\linewidth]{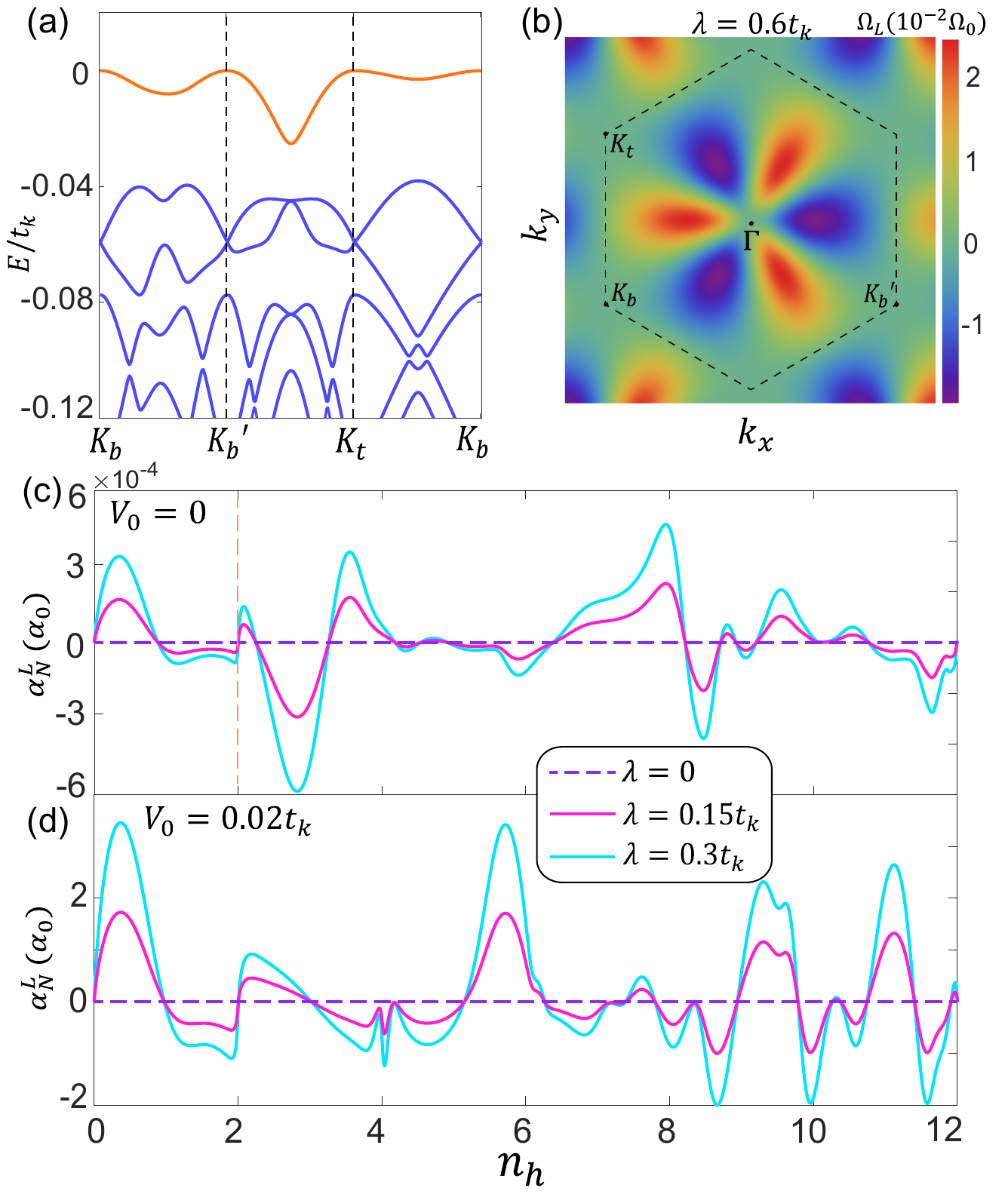}
		\caption{(a) Moir\'{e} bands of valley $\xi=+$ from the continuum model of Eq.~(\ref{eq:continummodel}) with the inter-layer tunneling strength $w=0.02t_k$ and $\theta=3^\circ$. (b) Momentum-space profile of $v(\bm{k)}\times v^L(\bm{k})$ [$\Omega_L(\bm{k})$] in units of $\Omega_0=a_0^2t_k^2/\hbar^2$ for the top moir\'{e} band in (a). In (a), (b) $\lambda=0.6 t_k$. (c) The Nernst coefficient $\alpha^L_{N}$ versus hole filling factor $n_h$ calculated from the continuum model under different trigonal warping strength $\lambda$. (d) $\alpha^L_{N}$ with intra-layer moir\'{e} potential $\delta V(\bm{r})$. The temperature is set as $k_B T=10^{-3}t_k$.}
		\label{fig:fig2}
\end{figure}

\emph{Minimal model with warping effect.}---Before diving into any realistic material systems, let us construct a minimal toy model for the twisted moir\'{e} bilayers. Without losing generality, one can merely consider a single band on each layer for simplicity. As illustrated in Fig.~\ref{fig:fig1}~(b), a twist angle $\theta$ between the two layers forms a moir\'{e} superlattice with an enlarged lattice constant $L_M=a_0/(2\sin\frac{\theta}{2})$, where $a_0$ denotes the original lattice constant. At small $\theta$, the moir\'{e} superlattice can be modeled by a continuum Hamiltonian $\mathcal{H}=\sum_{\xi}\int d\bm{r}\psi_{\xi}^\dagger(\bm{r})\mathcal{H}_{\xi}(\bm{r})\psi_{\xi}(\bm{r})$, where $\xi=\pm$ denotes the different valleys, and $\mathcal{H}_+$ ($\mathcal{H}_-$) for valley $K$ ($K^{\prime}$) is given by
\begin{equation}
\label{eq:continummodel}
		\mathcal{H}_+(\bm{r})=\begin{pmatrix}
			H_b(\bm{k})  & U(\bm{r})\\
			U^{\dagger}(\bm{r}) & H_t(\bm{k})
		\end{pmatrix}. 
\end{equation}
Each layer (indicated via $l=b/t$) is then governed by $H_l = -a_0^2t_k \tilde{\bm{k}}_l^2+a_0^3\lambda (\tilde{k}_{lx}^3-3\tilde{k}_{lx}\tilde{k}_{ly}^2)$, where $\tilde{\bm{k}}_l=\bm{k}-\bm{K}_l$, $t_k$ denotes the usual kinetic energy. A finite $\lambda$ introduces the trigonal warping effect, leaving the warped Fermi pockets around each Brillouin zone corner, as shown in Fig.~\ref{fig:fig1}~(b). The interlayer coupling in $\mathcal{H}_{\xi}$ is given by $U(\bm{r})=w[1+e^{-i\bm{G_2}\cdot\bm{r}}+e^{-i(\bm{G_1}+\bm{G_2})\cdot\bm{r}}]$, with $\bm{G_i}=4\pi/(\sqrt{3}L_M)(\cos 2(i-1)\pi/3,\sin 2(i-1)\pi/3)$ being the moir\'{e} reciprocal lattice vectors, and $w$ being the interlayer tunneling strength. The latter significantly affects the moir\'{e} energy bands, because of which, the top two valence bands exhibit an evident energy gap, as shown in Fig.~\ref{fig:fig2}~(a). 

In Fig.~\ref{fig:fig2}~(b) we plot the LVC $\Omega_L$ of the top moir\'{e} band, and the calculated LNE coefficient $\alpha_N^L$ is shown in Fig.~\ref{fig:fig2}~(c). Units $\Omega_0=a_0^2t_k^2/\hbar^2$ and $\alpha_0=e\tau k_Bt_k/\hbar^2$ are utilized here, and the temperature is set as $k_B T=10^{-3}t_k$. One may observe $\alpha_N^L=0$ when the warping strength $\lambda$ equals zero in Fig.~\ref{fig:fig2}~(c). This is because $\lambda = 0$ explicitly restores the intravalley $\mathcal{I}$ symmetry, rendering the LNE vanishes. More details are discussed in Supplementary Material~\cite{NoteX}.

We have also demonstrated that the presence of intra-layer moir\'{e} potential \cite{wu2019topological,li2021lattice}, does not alter our conclusion. To illustrate this, we consider a moir\'{e} potential described by $\delta V(\bm{r})=2 V_0\sum_{j=1}^3 \cos(\bm{G}_j\cdot \bm{r}+l\psi)$. The numerical results for $\alpha_N^L$ with $V_0=0.02 t_k$ and $\psi=91^\circ$ are shown in Fig.~\ref{fig:fig2}(d), and more related results are discussed in Supplemental Material~\cite{NoteX}. Based on the minimal model, an estimation of $\alpha^L_N \sim 0.03~\mu A/K$ can be made accordingly, utilizing parameters for MoS$_2$~\cite{kormanyos2013monolayer} with $a_0=3.2\AA$, $t\approx 0.86$ eV, $\lambda\approx 0.14$ eV, and $\tau\sim$1 ps.

\begin{figure}
		\centering
		\includegraphics[width=1.0\linewidth]{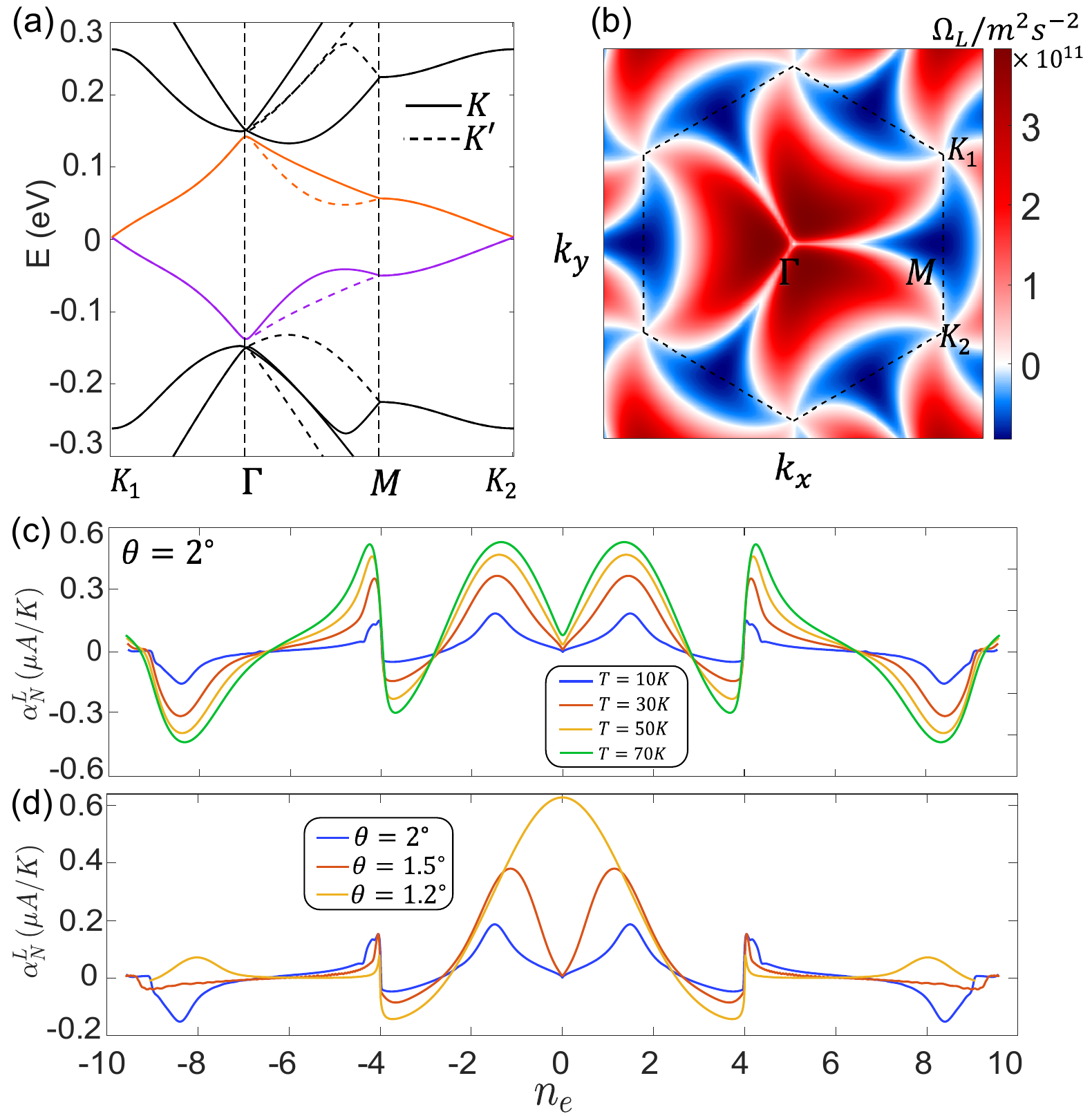}
		\caption{(a) The calculated moir\'{e} bands of TBG at $\theta=2^\circ$. The solid (dashed) lines denote that for $K$ and $K'$ valley, respectively, while the purple and orange lines represent the lowest conduction and highest valence bands, respectively. (b) momentum-space profile of layer velocity curvature $\Omega_L$ within the lowest conduction band (orange) in (a). (c) $\alpha^L_N$ as a function of filling factor $n_e$ for different temperatures $T$ at $\theta=2^\circ$. (d) $\alpha^L_N$ as a function of $n_e$ for different $\theta$ at $T$=10 K.}
		\label{fig:fig3}
	\end{figure}

\emph{LNE response in TBG.}---Now we apply the previously formalized LNE to TBG. For small twist angle $\theta$, the low-energy states of each layer in TBG can be effectively captured by a continuum model consisting of Dirac fermions, with an approximate Fermi velocity $\hbar v_F\approx 5.96$~eV$\cdot\text{\AA}$~\cite{bistritzer2011moire}. Additionally, the interlayer coupling in TBG introduces both the intra-sublattice tunneling $u_{AA}$ (AA stacking) and the inter-sublattice tunneling $u_{AB}$ (AB stacking). According to the Bistritzer-MacDonald model \cite{bistritzer2011moire}, these two tunneling strengths are considered to be $\sim 110~\mathrm{meV}$. As shown in Fig.~\ref{fig:fig3}~(a), we present the energy bands for both the $K$ and $K'$ valleys of TBG at $\theta=2^{\circ}$. Fig.~\ref{fig:fig3}(b) depicts the LVC $\Omega_L (\bm{k})$ for the lowest conduction band of $K$ valley, while the profile around $K'$ valley could be mapped out through the $\mathcal{T}$ symmetry operation. 

We investigate the properties of $\alpha^L_N$ for TBG by numerically solving the continuum model with an estimated scattering time $\tau \sim 1~\mathrm{ps}$ \cite{brida2013ultrafast}. Generally, the effective scattering time $\tau$ is estimated to be within $1\sim 10$ ps \cite{sharma2021carrier}. As shown in Fig.~\ref{fig:fig3}(c), $\alpha^L_N$ presents a great gate tunability. Evidently, $\alpha^L_N$ becomes larger with increasing temperatures, which is beneficial for experimental probing. We also show the twist angle dependence of $\alpha^L_N$ in Fig.~\ref{fig:fig3}(d). It can be observed that $\alpha^L_N$ exhibits more significant amplitudes at relatively smaller angles, attributed to the enhanced density of states there. Moreover, the two side peaks around $n_e=0$ merge together when $\theta$ approaches the magic angle (see $\theta=1.2^\circ$), leaving $\alpha^L_N\neq 0$ at the charge neural point. However, as shown in Fig.~\ref{fig:fig3}(c), we would again obtain $\alpha^L_N = 0$ at $n_e=0$ when $T\rightarrow 0$. In contrast to the layer Hall response~\cite{zhai2023time,NoteX}, LNE is even in the filling number, consistent with the relations given by Eq.~(\ref{eq:Mottrelation}). It is worth noting that $\alpha^L_N$ in TBG does not require the twist angles to be well-tuned, which greatly differs from the Berry curvature dipole's nonlinear transport effects in such systems~\cite{zhang2022giant}.

\begin{figure}
		\centering
		\includegraphics[width=1\linewidth]{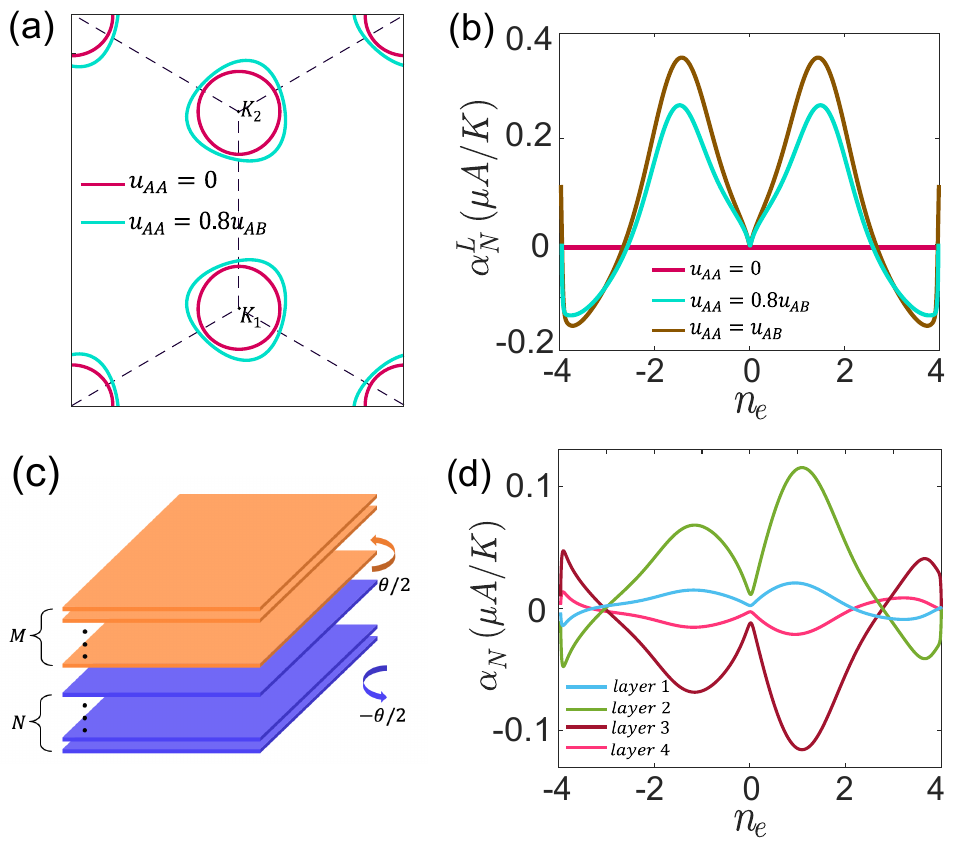}
		\caption{(a) The Fermi surface of TBG near $K$ points for $u_{AA}=0$ and $u_{AA}=0.8u_{AB}$ at $\theta=2^\circ$ and the Fermi energy $\mu=40$ meV. (b) The calculations of $\alpha_N^L$ for $u_{AA}=0, 0.8u_{AB}$ and $u_{AB}$. At $u_{AA}=0$, $\alpha_N^L=0$ because of the intravalley inversion symmetry. (c) Schematic structure of the twisted multilayer graphene system. (d) The gate dependence of $\alpha_N$ for each layer in TDBG at $\theta=2^\circ$. The temperature is set as $T=30$ K.}
		\label{fig:fig4}
\end{figure}

As discussed earlier, intravalley $\mathcal{I}$ symmetry breaking is necessary to support a finite LNE in moir\'{e} bilayers. This, here in TBG, is exactly realized by the finite interlayer tunneling, thus guaranteeing a finite LNE. As shown in Fig.~\ref{fig:fig4}(a), the Fermi surface features an obvious $\mathcal{I}$ symmetry with $u_{AA}=0$, which results in a zero LNE transport response. However, such symmetry is immediately broken by $u_{AA}\neq 0$, revealed by the trigonally warped Fermi surface; hence, a nonzero LNE transport response appears. The results discussed in Fig.~\ref{fig:fig3} exactly belong to this regime, and more numerical results of $\alpha^L_N$ with various $u_{AA}$ can be seen in Fig.~\ref{fig:fig4}~(b). 

Interestingly, what discussed above for LNE in TBG can be well captured by a pristine TBG model in the pseudo-Landau level representation \cite{liu2019pseudo}, which is written as
\begin{equation}
\label{eq:tbglandau}
H_{\xi}(\bm{k},\bm{r})=\hbar v_F[\bm{k}-\xi\frac{e}{\hbar}\bm{A}(\bm{r})\tau_z]\cdot\bm{\sigma}+3u_{AA}\tau_y.
\end{equation}
It can be easily recognized that, at the chiral limit ($u_{AA}=0$), TBG preserves intravalley $\mathcal{I}$ symmetry~\cite{tarnopolsky2019origin,wang2021chiral}, i.e., $I H_{\xi}(\bm{k},\bm{r}) I^\dagger=H_{\xi}(\bm{-k},\bm{-r})$ with the inversion operator $I=\tau_z \sigma_z$. Therefore, LNE is forced to vanish in TBG. This explains the numerical results shown in Fig.~\ref{fig:fig4}~(b). Moreover, one finds above an effective vector potential $\bm{A}(\bm{r})$, which provides a large layer-contrasted pseudomagnetic field in TBG [see Fig.\ref{fig:fig1}~(c)]. For example, a field of the order of magnitude $\sim 200$ T can be obtained at $\theta=2^\circ$. Microscopically, such an effective field can drive the carriers to transport along opposite directions in each layer (consistent with the picture constrained by symmetries) and give rise to the LNE response.

\emph{Application to twisted multilayer systems.}---Having studied the LNE in TBG, we now embark on other twisted multilayer systems. Generally, we consider a system in which $M$-layer graphene is stacked on top of another $ N$-layer graphene, forming a twist angle $\theta$, as depicted in Fig.~\ref{fig:fig4}(c). The layer Nernst coefficient can be similarly derived from the projected layer velocity $\hat{v}_L^i$ of the $i-$th layer with $\hat{v}_L^i=\{ \hat{v},\hat{P}^i \}/2$, where $\hat{P}^i$ is the projection operator onto the target $i\mathrm{th}$ layer. Under the overall $\mathcal{T}$ symmetry, the Nernst coefficients in each layer are explicitly constrained by $\sum_{i}\alpha_N^i=0$. 

We take the TDBG with $M=2$ and $N=2$ as an example. In Fig.\ref{fig:fig4}~(d), we illustrate the gate dependence of LNE coefficient $\alpha_N$ for each layer within TDBG at a twist angle of $\theta=2^\circ$.  Notably, prominent $\alpha_N$ values appear in the second and third layers, constituting the adjacent layers twisted with each other in TDBG. For the first and fourth layers, nonzero LNE responses can also be seen due to the presence of interlayer coupling in bilayer graphene. The observable asymmetry of $\alpha_N$ around the $n_e =0$ is ascribed to the absence of particle-hole symmetry. These results further verify our previous symmetry analysis and indicate that LNE is quite general in various layer structures.

\emph{Discussion and conclusion.}---We have introduced the LNE in several twisted moir\'{e} systems, along with the gate tunability, twisted angle and temperature dependencies. Remarkably, we find the twisted bilayers can support LNE with a very large transport coefficient. A closer comparison between the LNE coefficient found for TBG and other material systems studied to date is necessary. To the best of our knowledge, large Nernst responses with $\alpha_N \sim 10~$A/(K$\cdot$m) and $\alpha_N~\sim 15~$A/(K$\cdot$m) have been reported in magnetic Weyl semimetal $\mathrm{Co}_3\mathrm{Sn}_2\mathrm{S}_2$~\cite{ANE_Li_2020PRM} and $\mathrm{UCo}_{0.8}\mathrm{Ru}_{0.2}\mathrm{Al}$~\cite{ANE_Asaba_2021_SciAdv}, respectively. Another higher record is then held by the single crystal $\mathrm{MnBi}$ with $\alpha_N \sim 45~$A/(K$\cdot$m)~\cite{ANE_He_2021_Joule}. For TBG considered in this work, $\alpha^L_N \approx 0.5~\mu$A/K can be obtained at a moderate temperature $T=70$ K (roughly same $T$ as in the above materials) and an easily achieved twisted angle $\theta = 2^\circ$. When considering the thickness about $0.5$ nm of bilayer graphene, such response coefficient is converted to be $\sim 10^3~$A/(K$\cdot$m).

\emph{Acknowledgements.}---This work is supported by the NSF of China (Grant No. 12104043), the National Key R\&D Program of China (Grant No. 2020YFA0308800), and the Strategic Priority Research Program of the Chinese Academy of Sciences (Grant No. XDB30000000). The authors appreciate the helpful discussions with Cong Xiao and Junxi Duan.


%

\end{document}